\documentclass[a4paper,11pt]{article}
\usepackage{jcappub}
\usepackage{graphicx}
\usepackage{amssymb}
\usepackage{bm}
\usepackage{color}

\newcommand{\nab}{\mbox{\boldmath $\nabla$} {}}
\newcommand{\be}{\begin{equation}}
\newcommand{\ee}{\end{equation}}
\newcommand{\gsim}{\gtrsim}
\newcommand{\lsim}{\lesssim}
\newcommand{\bea}{\begin{eqnarray}}
\newcommand{\eea}{\end{eqnarray}}
\newcommand{\bean}{\begin{eqnarray*}}
\newcommand{\eean}{\end{eqnarray*}}

\newcommand{\G}{\,{\rm G}}
\newcommand{\Mpc}{\,{\rm Mpc}}
\newcommand{\Gpc}{\,{\rm Gpc}}
\newcommand{\GeV}{\,{\rm GeV}}
\newcommand{\const}{{\rm const}}
\newcommand{\Fig}[1]{Fig.~\ref{#1}}
\newcommand{\yjfm}[3]{, J.\ Fluid Mech.\ {\bf #2}, #3 (#1).}

\definecolor{dgreen}{rgb}{0,0.7,0} % darkgreen

\def\tauA{\tau_{\rm A}}
\def\xiM{\xi_{\rm M}}
\def\kM{k_{\rm M}}
\def\cs{c_{\rm s}}
\def\uu{{\bf u}}

\def\BB{{\bf B}}
\def\JJ{{\bf J}}
\def\AAA{{\bf A}}
\def\urms{u_{\rm rms}}
\def\half{{\textstyle{1\over2}}}
\def\onethird{{\textstyle{1\over3}}}
\def\la{{\lambda}}
\newcommand{\SSSS}{\mbox{\boldmath ${\sf S}$} {}}

\newcommand{\bk}{{\bf k}}
\newcommand{\bde}{{\bf e}}
\newcommand{\ra}{\rightarrow}

\title{Scale-invariant helical magnetic field evolution and the duration of inflation}

\author[a,b,c,1]{Tina Kahniashvili,
}
\author[d,e,f,g]{Axel Brandenburg,}
\author[h]{Ruth~Durrer,}
\author[i,c]{Alexander G.\ Tevzadze,}
\author[j,a]{and Winston Yin}

\affiliation[a]{McWilliams Center for
Cosmology and Department of Physics, Carnegie Mellon University,
 Pittsburgh, PA 15213, USA}
\affiliation[b]{Department of Physics, Laurentian University, Ramsey
Lake Road, Sudbury, ON P3E 2C, Canada}
\affiliation[c]{Abastumani Astrophysical Observatory, Ilia State University,
 0194 Tbilisi, Georgia}
\affiliation[d]{Laboratory for Atmospheric and Space Physics, University of Colorado, Boulder, CO 80303, USA}
\affiliation[e]{JILA and Department of Astrophysical and Planetary Sciences, University of Colorado, Boulder, CO 80303, USA}
\affiliation[f]{Nordita, KTH Royal Institute of Technology and Stockholm University, 10691 Stockholm, Sweden}
\affiliation[g]{Department of Astronomy, AlbaNova University Center,
Stockholm University, 10691 Stockholm, Sweden}
\affiliation[h]{D\'epartement de Physique
Th\'eorique and Center for Astroparticle Physics, Universit\'e de
Geneve,  1211 Gen\'eve 4, Switzerland}
\affiliation[i]{Faculty of Exact and Natural
Sciences, Javakhishvili Tbilisi State University,  Tbilisi, 0179, Georgia}
\affiliation[j]{Department of Physics, University of California Berkeley, Berkeley, CA 94720, USA}

\emailAdd{tinatin@andrew.cmu.edu}
\abstract{
We consider a scale-invariant helical magnetic
field generated during inflation.
We show that, if the mean magnetic helicity density of such a field
is measured, it can be used to determine a lower bound on the duration
of inflation.
Upper bounds can be used to derive constraints on the minimal duration
of inflation if one assumes that the magnetic field generated during
inflation is helical.
Using three-dimensional simulations, we show that an initially scale-invariant field develops, which is similar both with and without
magnetic helicity. In the fully helical case, however, the magnetic field appears to
have a more pronounced folded structure.
}

\begin{document}
\maketitle
\flushbottom

\section{Introduction} \label{sec:intro}
Many astrophysical observations indicate that coherent
magnetic fields of the order of microGauss are present in galaxies and
clusters \cite{Widrow:2002ud}.
There is also evidence that fields of more than $10^{-17}\G$
with large correlation lengths permeate the entire universe,
even the voids \cite{Neronov:1900zz,Taylor:2011aa}.
Although the origin of these fields
is under debate, it is assumed that the observed fields originated from cosmological or astrophysical seed magnetic fields and were amplified
during structure formation, either via adiabatic compression or
magnetohydrodynamic (MHD) turbulent dynamo instabilities
\cite{Kulsrud:2007an,Kandus:2010nw,Durrer:2013pga}. The statistical
properties of the resulting magnetic field (its amplitude, spectral
shape, correlation length, etc) strongly depend on the initial
conditions, i.e.\ the generation mechanisms. In the case of magnetic
fields generated through causal processes (all astrophysical
scenarios as well as primordial magnetogenesis occurring in the early universe, but {\em after} inflation), the correlation length is strictly limited by the Hubble horizon scale at the moment of magnetic field generation.
Accounting for the free decay of the magnetic field during the
expansion of the universe, it may reach the scale of galaxies, but not much more \cite{Kahniashvili:2012uj}.
Such fields are not expected to be correlated on Mpc scales.
This limitation does not apply when considering seed magnetic
fields generated during inflation. In this case the magnetic field is
generated by the amplification of quantum fluctuations, and its
correlation length can be very large.

The evolution of magnetic fields during the expansion of the universe as
well as their observable signatures strongly depend on the helicity of the
initial seed field~\cite{Banerjee:2004}.
Magnetic helicity is observed in a number of astrophysical
objects ranging from stellar outflows \cite{Ching:2016}
to jets from active galactic nuclei (AGNs) \cite{Ensslin:2002gn}.
While these are all examples of astrophysically generated helical
fields, there might now also be evidence of helical intergalactic
fields from the gamma-ray arrival directions observed by Fermi-LAT
\cite{Tashiro:2015,Chen:2015}.
Even if the initial helicity is not close to its maximum
possible value, the fractional helicity grows during the evolution
through MHD turbulence \cite{BM99}, leading to a maximally helical
configuration of the observed fields at late time \cite{Tevzadze:2012kk}.

Primordial magnetic helicity, if detected, will be a direct
indication of parity (mirror symmetry) violation in the early universe, and may be related to the matter-antimatter asymmetry
problem \cite{Long:2013tha,Fujita:2016igl}. Generation of a helical magnetic field
in the early universe obviously requires a parity
violating source, which can be present during cosmological phase
transitions (electroweak or QCD) or during inflation
\cite{Cornwall:1997ms,Giovannini:1997eg,Field:1998hi,Vachaspati:2001nb,
Tashiro:2012mf,Sigl:2002kt,Subramanian:2004uf,Campanelli:2005ye,Semikoz:2004rr,
DiazGil:2007dy,Campanelli:2008kh,Campanelli:2013mea,Jain:2012jy,
Boyarsky:2011uy,
Calzetta:2014eaa,Caprini:2014mja,Piratova:2014mra}\footnote{See
Ref.~\cite{Wagstaff:2014fla} for questioning the efficiency of the
cosmological phase transition originated helical magnetogenesis.}.

One of the main observables helping to constrain primordial helical magnetic
fields are parity-odd CMB cross correlations (such as temperature --
$B$-polarization, as well as $E$- and $B$-polarization) which are absent
in the standard cosmological scenario, as well as in the models with
non-helical magnetic fields
\cite{Pogosian:2001np,Caprini:2003vc,Kahniashvili:2005xe,Kunze:2011bp,
Kahniashvili:2014dfa,Ballardini:2014jta}.
The amplitude of parity-odd correlations in the  CMB depends both on the
magnetic field amplitude and on its helicity. At this point it is
important to note that Faraday rotation of the CMB polarization by the
magnetic field~\cite{Kosowsky:1996yc} is insensitive to magnetic
helicity
\cite{Ensslin:2003ez,Campanelli:2004pm,Kosowsky:2004zh,
Kahniashvili:2008hx,Ade:2015cao},
allowing in this way to limit (or detect) the amplitude of the magnetic
field by Faraday rotation measurements; Ref.~\cite{Ade:2015cva} finds
upper bounds on the magnetic field from CMB data of the order of a few
$10^{-9}\G$ on the scale of $1\Mpc$.

A helical magnetic field is a natural source of parity-odd CMB fluctuations. These can be induced
by different and even more exotic means like a generic CPT violation
\cite{Cabella:2007br,Xia:2007qs,Feng:2006dp,Xia:2012ck,
Xia:2009ah,Li:2009rt,
Li:2008tma,Gruppuso:2011ci,Gubitosi:2012rg,Li:2013vga,
Gluscevic:2012me,Ben-David:2014mea,Tasson:2014dfa,Kaufman:2014rpa},
a Chern-Simons coupling of the electromagnetic field
\cite{Lue:1998mq,Xia:2008si,Saito:2007kt},
 a homogeneous magnetic field
\cite{Scannapieco:1997mt,Scoccola:2004ke,Demianski:2007fz,
Kristiansen:2008tx,Adamek:2011pr},
 Lorentz symmetry breaking
\cite{Carroll:1989vb,Kostelecky:2007zz,Cai:2009uc,Ni:2007ar,Casana:2008ry,Caldwell:2011pu,
MosqueraCuesta:2011tz,Kamionkowski:2010rb,Gluscevic:2010vv,Mewes:2012sm,
Ni:2009qm,Miller:2009pt,Ni:2009gz}, or a non-trivial cosmological topology \cite{Lim:2004js,Carroll:2004ai,Alexander:2006mt,Satoh:2007gn}. If some non-vanishing
parity-odd CMB correlations will be detected, the corresponding
angular power spectrum and the frequency spectrum must be measured
sufficiently precisely in order to distinguish between the different
possibilities \cite{Gruppuso:2015xza}.

In this paper we focus on magnetic helicity generated during inflation.
We show that, if inflation generates a scale-invariant helical
magnetic field, this can be used to constrain the largest scale
amplified during inflation, which characterizes the total duration of
inflation, not only its duration after Hubble exit which is well known
to be of the order of 50 to 60 $e$-folds. Even if inflation-generated
magnetic fields are not necessarily helical, this is still an exciting
prospect.
To our knowledge, this would be the first observation which could,
at least in principle, access the beginning of inflation, even it the
corresponding scale today is much larger than the present Hubble scale.

The current limits on parity-odd fluctuations in the CMB are already
able to constrain the correlation length of a possible helical magnetic
field, which is well beyond the present Hubble scale.
CMB data can constrain only inflation-generated, scale-invariant
helical magnetic fields \cite{Kahniashvili:2014dfa}.
The magnetic field generated
during inflation induces {\it causal} modes of density perturbations with
a finite correlation length. In particular, the magnetic sources for all
modes (scalar, vector, and tensor) are determined by the energy-momentum
tensor of the magnetic field and are given by convolutions of the magnetic
field \cite{Caprini:2003vc,Kahniashvili:2006hy,Kahniashvili:2005xe}.

We also investigate numerically the velocity field generated by the magnetic
field (including both vorticity and longitudinal velocity).
We show that magnetic
fields generated during inflation induce a {\it causal} velocity field
with a white-noise spectrum at large scale. An important difference in
our simulation setup relative to those of previous studies
(see Ref.~\cite{Subramanian:2015lua} for a review) is the treatment
of the backreaction of fluid perturbations onto the magnetic field.
We show that, for wavenumbers that are within the Hubble horizon,
an initially scale-invariant spectrum quickly develops
a turbulent forward cascade with a Kolmogorov-like spectrum.
Furthermore, when the fractional helicity is initially below unity,
it grows slowly and would eventually reach unity.

The outline of the paper is as follows: in Sec.~II we present the main
statistical characteristics of a helical magnetic field.
In Sec.~III we discuss existing limits on magnetic
helicity from CMB data. We also show the evolution of primordial helical magnetic fields. We discuss future experimental prospects in Sec.~\ref{s:prospects}
and conclude in Sec.~\ref{s:con}.

We work with comoving quantities (magnetic field, length scales,
wave numbers etc), where the scale factor is normalized to unity today,
$a(t)=(1+z)^{-1}$, and $z$ is the redshift. We employ natural units ($\hbar = c = k_B= 1$) along with Lorentz-Heaviside units for the magnetic field, so there are no factors of $4\pi$ in the
Maxwell equations and the magnetic energy density is $B^2/2$.

\section{Modeling a Turbulent Helical Magnetic Field}

We assume that a cosmological helical magnetic field was generated
during inflation with a scale-invariant spectrum on scales below the
horizon scale at the time of generation. There are of course also other possibilities leading to blue helical magnetic fields from inflation~\cite{Anber:2006xt,Durrer:2010mq}
After inflation, the field is correlated over super-Hubble scales.
The energy density of the magnetic field
must be small enough in order to preserve the isotropy of the
universe, and not to spoil the inflationary stage
\cite{Bonvin:2011dt}\footnote{There are regions in model parameter
space where the magnetic (or electric) field fluctuations during inflation
are so large that they backreact on the inflationary expansion and also
invalidate the linear perturbation assumption. We assume that this does not
occur, which is compatible with the generation of scale-invariant
seed fields which are strong enough to induce the large-scale magnetic field
observed today~\cite{Ratra:1992ab,Ratra:1991ab,Demozzi:2009ab,
Kanno:2009ab,Emami:2009vd,Motta:2012rn,Demozzi:2012wh}.
Backreaction is not relevant for the models considered here.}.
In what follows, we assume that the magnetic energy density
is a first-order perturbation on the standard
homogeneous and isotropic background cosmological model.

\subsection{Statistical properties of helical magnetic fields}
The plasma generated during reheating after inflation is highly conductive
and can be treated in the MHD limit.
If the magnetic-field were just frozen-in, the
spatial and temporal dependence of the field decouple due to flux conservation, ${\bf B}({\bf x}, t) \propto {\bf B_0}({\bf x})/a^2(t)$, where ${\bf x}$ is the position vector, $t$ is the conformal time with
$dt=d\tau/a(t)$ (with $\tau$ denoting the physical time), and $a(t)$
is the cosmological scale factor.
However, the evolution of a
primordial magnetic field is a complex process influenced by MHD as
well as the dynamics of the universe. In previous work
\cite{Kahniashvili:2012vt}, we studied nonhelical inflation-generated
magnetic field evolution during the expansion of the universe,
in particular during cosmological phase transitions. Below, we
present a similar study, but for helical fields (the evolution of helical magnetic fields for
different initial spectra is presented in Ref.~\cite{Brandenburg:2016odr}.
the magnetic energy
density is given as ${\mathcal E}_{\rm M} = \langle {\bf B({\bf x})}^2 \rangle/2$, where
$\langle...\rangle$ denotes the average over all space\footnote{In
what follows we omit $\langle ... \rangle$ for simplicity when determining
{\it mean} energy densities.}. The kinetic energy density is written as
${\mathcal E}_{\rm K} = \langle \rho {\bf u({\bf x})}^2 \rangle/2$.

The mean helicity density of the magnetic field in a volume $V$ is given by
\begin{eqnarray}
{\mathcal H}_{\rm M} &=&\frac{1}{V} \int_V d^3{\bf x} ~{\bf A({\bf x})}\cdot{\bf B({\bf x})}, \label{helicity_def}
\end{eqnarray}
with ${\bf B}=\nab\times{\bf A}$ and $\bf A$ being the vector potential.
In principle, this integral is gauge-invariant only if the magnetic field vanishes on the boundary or toward infinity.
However, the magnetic helicity is also gauge-invariant for periodic
systems with zero net magnetic flux, as shown in Ref.~\cite{gauge}. We assume that the universe can be well approximated by a domain with periodic boundary conditions, provided the dimension of the domain is large compared to the Hubble scale today.
In this case, the mean magnetic helicity density, ${\mathcal H}_{\rm M}$, is a well-defined quantity.

Assuming that the magnetic field is a Gaussian random field,
its two-point correlation function in wavenumber space is given by
\begin{equation}
\langle B_i^* ({\bf k}) B_j({\bf k'}) \rangle =
(2\pi)^3 \delta({\bf k} -{\bf k'}) \, {\mathcal F}_{ij}\!({\bf k})\,,
\label{2-point}
\end{equation}
where $\delta^{3}({\bf k}-{\bf k'})$ is the
three-dimensional Dirac delta function.
The most general ansatz
for ${\cal F}_{ij}({\bf k})$ satisfying statistical isotropy,
${\cal F}_{ij}({\bf k}) =  {\cal F}_{ij}({ k})$ with $k=|{\bf k}|$,
as well as the divergence-free condition, $\nab \cdot {\mathbf B}=0$,
is of the form
\begin{equation} \frac{
{\mathcal F}_{ij}({\bf k})}{(2\pi)^3}= (\delta_{ij}-{\hat k}_i {\hat k}_j)
\frac{E_{\rm M}(k)}{4\pi k^2} + i \epsilon_{ijl} {k_l}
\frac{H_{\rm M}(k)}{8\pi k^2}. \label{eq:4.1}
\end{equation}
Here, $\hat{k}_i=k_i/k$ are the components of the unit wavevector,
$\delta_{ij}$ is the Kronecker delta and
$\epsilon_{ijl}$ is the antisymmetric tensor.
$E_{\rm M}(k)$ and $H_{\rm M}(k)$ are respectively
the spectral energy and helicity densities of the magnetic field.
We use the Fourier transform convention $F_j({\bf k}) = \int d^3x \,
e^{i{\bf k}\cdot {\bf x}} F_j({\bf x})$,
so the spectral tensor,
${\mathcal F}_{ij}({\bf k}) = \int d^3{\bf r} \,
e^{i{\bf k}\cdot{\bf r}} {\mathcal F}_{ij}({\bf r})$ is the Fourier transform of the magnetic two-point correlation function
${\cal F}_{ij} ({\bf r}) = \langle B_i({\bf x})B_j({\bf x} + {\bf r})\rangle$.

With the above notations, the mean magnetic energy density is
${\mathcal E}_{\rm M} =\delta_{ij} \lim_{{\bf r} \rightarrow 0}
 {\cal F}_{ij} ({\bf r})/2$, and it is
related to $E_{\rm M}(k)$ through
\begin{equation} \label{e:EB}
{\mathcal E}_{\rm M} = \int_0^{\infty} dk \, E_{\rm M}(k).
\end{equation}
Similarly, the mean magnetic helicity density
${\mathcal H}_{\rm M}$ is related to $H_{\rm M}(k)$ through
\begin{equation}
{\mathcal H}_{\rm M} = \int_0^{\infty} dk \, H_{\rm M}(k) \, .
\end{equation}
Here $E_{\rm M}(k)$ and $H_{\rm M}(k)$ are related to
the symmetric and antisymmetric magnetic power spectra,
$P_{\rm M}(k)$ and $P_{\rm H}$, respectively, through\footnote{The
spectral correlation function is defined as \cite{Caprini:2003vc},
\begin{equation}
\langle B_i^* ({\bf k}) B_j({\bf k'}) \rangle =
(2\pi)^3 \delta({\bf k} -{\bf k'}) \,
\Big[(\delta_{ij}-{\hat k}_i {\hat k}_j) P_{\rm M}(k)
+ i\varepsilon_{ijl} {\hat k}_l P_{\rm H}(k)\Big]\, .
\label{2-point-2}
\end{equation}
}
\begin{equation}
E_{\rm M}(k) = \frac{k^2 P_{\rm M}(k)}{2\pi^2}, \quad
H_{\rm M}(k) = \frac{k P_{\rm H}(k)}{\pi^2}.
\end{equation}
Here $P_{\rm M}(k)$ is symmetric under parity
transformation, $\bk\ra-\bk$, while $P_H(k)$ is antisymmetric.
In the following, we also consider the
magnetic field correlation length defined as
\begin{equation}
\xi_{\rm M} = \frac{1}{\mathcal E}_{\rm M}\int_0^{\infty} dk \,
k^{-1} E_{\rm M}(k),
\label{xi_def}
\end{equation}
which is fully determined by the spectral energy density
of the magnetic field.

The Cauchy--Schwarz inequality for the magnetic field -- the
realizability condition -- reads (see also \cite{bdkmty17})
\begin{equation}
|{\mathcal H}_{\rm M}| \leq 2 \xi_{\rm M} {\mathcal E}_{\rm M} \, .
\label{cs-inequality}
\end{equation}
The spectral form of the realizability condition is $k|H_{\rm M}(k)|\leq 2 E_{\rm M}(k)$, i.e., $|P_H(k)| \leq P_{\rm M}(k)$, and equality is reached only
in the maximally helical case, i.e.\ when the magnetic field is fully
right-handed or fully left-handed\footnote{
It is instructive to express
the Fourier transform of the magnetic field in a helicity basis. Choosing
$\bde_1(\bk),\bde_2(\bk)$ such that $(\bde_1(\bk),\bde_2(\bk),\hat\bk)$
form a right handed orthonormal system, we introduce \bea \bde_\pm
&=&\frac{1}{\sqrt{2}}(\bde_1\pm i\bde_2)  \quad \mbox{and }\\{\bf B}
&=& B_+\bde_+ + B_-\bde_- \,.\eea Introducing this decomposition we obtain
\bea \langle B^*_+({\bf k})B_+({\bf k'}) + B^*_-({\bf
k})B_-({\bf k'}) \rangle = \qquad  \qquad && \nonumber \\ \qquad
(2\pi)^3 \delta^{3} ({\bf k}-{\bf k'})2P_{\rm M}(k)   && \label{ef:PB} \\ \langle
B^*_+({\bf k})B_+({\bf k'}) - B^*_-({\mathbf k})B_-({\bf
k'}) \rangle =  \qquad   \qquad  && \nonumber   \\ \qquad  (2\pi)^3
\delta^{3} ({\bf k}-{\bf k'})2P_H(k) \,. &&  \label{ef:PH}
\eea
Adding and subtracting (\ref{ef:PB}) and (\ref{ef:PH}) we find
$P_{\rm M}+P_H\geq 0$ and $P_{\rm M}-P_H\geq 0$.
This implies that $P_{\rm M}\ge |P_H|$.}.
Assuming that the symmetric and antisymmetric power
spectra of the magnetic field are given by simple power laws,
\bea
P_{\rm M}(k) \propto k^{n_{\rm M}} \quad\mbox{and}\quad P_{\rm H}(k) \propto k^{n_H}
\eea
for $k_{\min}\leq k\leq k_{\max}$, and zero otherwise,
the constraint on their relative amplitudes implies $n_H \ge
n_{\rm M}$, see~\cite{Durrer:2003ja}\footnote{Some confusion from the spectral
form of the realizability might occur because the formulation above uses
the same spectral indices for the whole
spectrum, while in reality the spectral shapes are different in the
long-wave and inertial regimes \cite{bdkmty17}.}.

Finally, we introduce the fractional helicity $\sigma$ with $|\sigma|\leq 1$ as
\begin{equation}
\sigma  =\frac{|{\mathcal H}_{\rm M}|}{2\xi_{\rm M} {\mathcal E}_{\rm M}}\,.
\end{equation}
For fully helical magnetic fields with known mean helicity density
${\mathcal H}_{\rm M}$, the correlation length is determined through
$2 \xi_{\rm M} \equiv|{\mathcal H}_{\rm M}|/{\mathcal E}_{\rm M}$, while for an
arbitrary helical field we have a lower limit for the correlation length, i.e.\ $2 \xi_{\rm M} >|{\mathcal H}_{\rm M}|/{\mathcal E}_{\rm M}$,
or, in terms of $\sigma$,
$ 2 \xi_{\rm M} =|{\mathcal H}_{\rm M}|/(\sigma {\mathcal E}_{\rm M})$.

We assume that the magnetic field generated during inflation
results in a scale-invariant spectrum $n_{\rm M} \rightarrow -3$.
Note that, in contrast to previous considerations,
e.g., Ref.~\cite{Subramanian:2015lua} and references therein,
there is a lower cutoff wavenumber for the scale-invariant spectrum,
$k_{\min}$, below which the spectral magnetic energy
and helicity densities vanish.
This is introduced to prevent $\xiM$ from diverging.
However, the effective $k_{\min}$ is really determined by
the physics of inflation and corresponds to the largest length scale
$k_{\min}^{-1}$ of the magnetic quantum-mechanical fluctuations generated
during inflation, above which the initial magnetic two-point
correlation function vanishes or has a sharp cut-off.
This implies that the real space two-point correlation function obeys
${\mathcal F}_{ij}({\bf r}) \rightarrow 0$ for $|{\bf r}| > k_{\min}^{-1}$,
so for $k<k_{\min}$ the spectral magnetic energy density scales as
$E_{\rm M}(k) \propto k^\alpha$ \cite{monin} with
$\alpha \geq 4$ \cite{Durrer:2003ja}, as required by the causality
and divergence-free conditions for the magnetic field.
The realizability condition then implies $H_{\rm M}(k) \propto k^\beta$
with $\beta \geq 3$ \cite{Caprini:2003vc}.
\subsection{Magnetic field correlation length}

\begin{figure}[t!]\begin{center}
\includegraphics[width=.7\columnwidth]{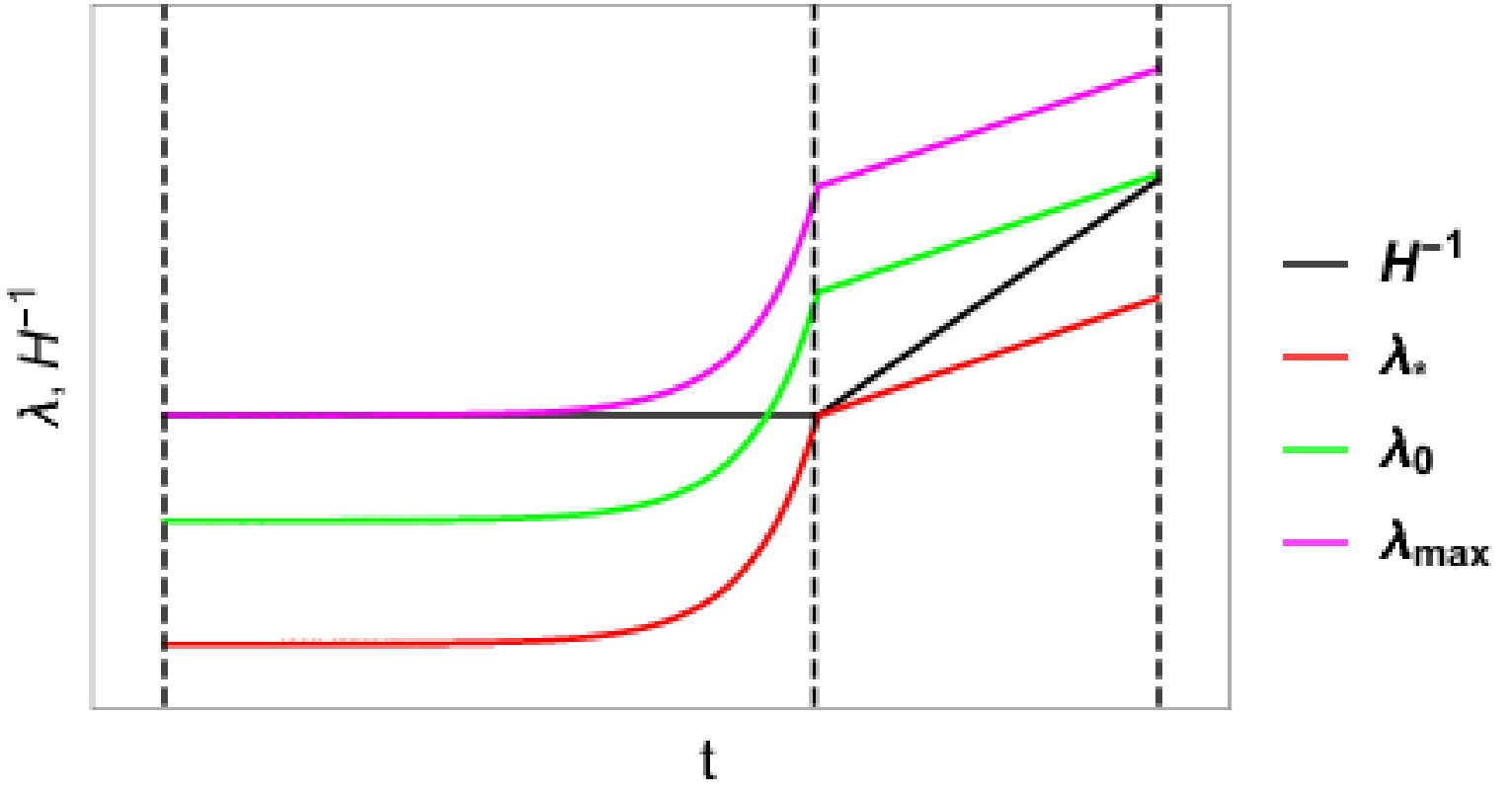}
\end{center}\caption{\label{f:infla} Different scales exit the horizon
$H^{-1}$ at different times. We indicate the largest amplified scale,
$\la_{\max}= a/k_{\min}$, the present Hubble scale, $\la_0=a/H_0$ and
the smallest amplified scale $\la_*$. The dashed vertical lines indicate
(from left to right) the beginning of inflation, the end of inflation,
and the present time.}
\end{figure}

Let us first consider the maximally helical case.
Measuring both the magnetic energy and helicity densities,
we can infer the largest length scale in the system as
\be
k_{\min}^{-1} = {\xi_{\rm M}} \simeq
\left(\frac{|{\mathcal H}_{\rm M}|}{2{\mathcal E}_{\rm M}}\right)_{\rm f}\, .
\label{kmin1}
\ee
Here, the suffix f indicates the end of inflation.
The only natural infrared cutoff for inflation is the horizon scale at
the beginning of inflation. Length scales larger than this are super horizon
already at the beginning of inflation and are therefore never amplified.
If a significant value of magnetic helicity were detected,
and if the spectrum of the field is indeed compatible with a
scale-invariant one, this would determine the cutoff wavenumber $k_{\rm min}$,
the length scale that exits the horizon at the beginning of inflation.
After some $e$-folds, also the present Hubble scale exits.
Their ratio determines the number of $e$-folds of inflation {\em before}
Hubble exit,
\be
\log(H_0/k_{\min}) = N_H\,.
\ee
The number of e-folds of inflation after Hubble exit, let us call it $N_e$
is rather well known, $N_e\sim 55$--$60$ \cite{Ade:2015lrj}, and therefore
this allows us to determine the duration of inflation. This situation
is illustrated in Fig.~\ref{f:infla}. {\it To our knowledge, this is the first time that the
possibility is proposed to determine the full duration of inflation.}

This rather simple but fascinating observation is our main result:
it implies that { a maximally helical scale-invariant magnetic field
generated during inflation carries information about the horizon scale
at the beginning of inflation.} A measurement of ${\cal E}_{\rm M}$ and
${\cal H}_{\rm M}$ allows us to retrieve this information;
see Appendix~\ref{Monochromatic}.

In (\ref{kmin1}), ${\cal E}_{\rm M}$ and ${\cal H}_{\rm M}$ are the
values soon after inflation, once the magnetic field is fully helical.
Let us now discuss how this ratio changes by subsequent MHD processes
within the Hubble horizon.
It is well known \cite{FPLM75,Brandenburg:2005xc} that the correlation length increases
subsequently during the magnetic field decay and that the speed
is faster for maximally helical fields. This growth process (within the Hubble horizon) starts
shortly after inflation and lasts until recombination.
Denoting the reheating temperature after inflation by $T_{\rm f}$, the growth
factor for the correlation length for maximally helical fields
through the inverse cascade is $(T_{\rm f}/T_{\rm rec})^{p}$.
For fully helical fields, one has $p=2/3$,
introducing a high inflation scale with
for $T_{\rm f} \sim 10^{15}\GeV$ and using $T_{\rm rec} \simeq 0.25\,$eV,
the correlation length grows by a factor of the order of $10^{16}$.
However, as we shall discuss in the next section, for a scale-invariant
magnetic field without subinertial range, we find $p\simeq 0.2\ll 2/3$
and the growth of the correlation length is reduced to a factor $10^{5}$.
No theory for the exponent 0.2 is known, and it may not be
a universal one, but this is secondary for our present purpose,
because uncertainties in the exponent would only yield an additional
correction when
inferring $k_{\min}$ by determining ${\mathcal H}_{\rm M}$ and
${\mathcal E}_{\rm M}$ through their effects on the CMB.
More precisely,
\be
k^{-1}_{\min}\simeq \left(\frac{|{\mathcal H}_{\rm M}|}{2{\mathcal E}_{\rm M}}
\right)_{\rm f} \sim (T_{\rm f}/T_{\rm rec})^{p} \,
\frac{|{\mathcal H}_{\rm M}|}{2 {\mathcal E}_{\rm M}}\,.
\ee

In the case of partially helical fields, there is an additional step.
The {\it super-horizon} modes retain the initial conditions while the
{\it sub-horizon} modes are influenced by MHD processes and fields
on sub-horizon scales rapidly become maximally helical through magnetic
decay while magnetic helicity is conserved.
Thus, the ratio $|{\cal H}_{\rm M}|/2{\mathcal E}_{\rm M}$ increases until
a maximally helical case ($\sigma=1$) is reached.
In fact, the MHD processes lead to re-distribution of helicity at large
scales and the fractional helicity is a time dependent (growing) function.
In this situation, the above equality becomes only a limit,
which, in terms of $T_{\rm rec}/T_{\rm f}$, is
\be
(T_{\rm rec}/T_{\rm f})^{p} > k_{\min} |{\mathcal H}_{\rm M}|/2
{\mathcal E}_{\rm M}.
\ee
Thus, for partially helical magnetic fields, we only have $|{\cal H}_{\rm M}|
< {\cal H}_{\rm M}^{\max}\equiv 2\xi_{\rm M}{\cal E}_{\rm M}$.

\section{Evolution and realizability condition for scale-invariant
helical magnetic fields}

We now present and discuss  results from numerical
simulations of the inverse cascade of helical magnetic fields.
Numerical simulations show that the well-known inverse cascade of
magnetic helicity \cite{Pouquet:1976zz,Christensson:2000sp} is strongly
reduced if the initial magnetic seed field has a scale-invariant spectrum
\cite{Brandenburg:2016odr}.
Specifically, we have ${\cal E}_{\rm M}\propto t^p$ and
$\xi_{\rm M}\propto t^q$ with $p\approx q\approx0.2$ instead of
$p=q=2/3$, so the correlation length grows much more slowly
than for causal fields with $n_{\rm M}=2$. This is not surprising; the smaller the $n_{\rm M}$, the more of the magnetic energy is
contained in super-horizon modes which are not affected by plasma processes.

Current limits on the primordial magnetic field from
CMB and large-scale structure data are around 1--2\,nG;
see~\cite{Kahniashvili:2012dy,Yamazaki:2013hda,Ade:2013zuv} and
references therein. The realizability condition then implies
${\mathcal H}_{\rm M} \leq 0.1 \, {\rm nG}^2 \xi_{\rm M}$, and
equality is reached for a maximally helical field.
As we have already noted above, the correlation length of
causally (post-inflation) generated magnetic fields is
limited by the Hubble scale $\lambda_H$ at the moment of
generation,
\begin{equation}
\lambda_{H} = 5.8 \times 10^{-10}~{\rm Mpc}\left(\frac{100\GeV}
{T_\star}\right) \left(\frac{100}{g_\star}\right)^{{1}/{6}},
\label{lambda-max-H}
\end{equation}
where $T_\star$ and $g_\star$ are the temperature and number of
relativistic degrees of freedom at the generation moment $t_\star$
\cite{Kahniashvili:2012uj}.
Accounting for the inverse cascade for maximally helical fields
[which increase the correlation length as
$(t_{\rm rec}/t_\star)^{2/3}$, i.e. during the radiation dominated epoch
the correlation length increases by $(T_\star/0.25~{\rm eV})^{2/3}$],
the maximal value of the correlation length for the causal field is given by
\begin{equation}
\lambda_{\rm max} \approx 10^{-2}~{\rm Mpc}\left(\frac{100\GeV}
{T_\star}\right)^{1/3} \left(\frac{100}{g_\star}\right)^{{1}/{6}},
\label{lambda-max}
\end{equation}
which is substantially smaller than the correlation length of the magnetic
field generated during inflation (see Ref.~\cite{Sharma:2017eps} for a
recent study on inflationary magnetogenesis).

We construct a random initial magnetic field in Fourier space as
\begin{equation}
B_i({\bf k})=B_0\left(\delta_{ij}-{\hat k}_i {\hat k}_j+i\tilde\sigma\epsilon_{ijl} {\hat k}_l\right)
f_j({\bf k})\, |{\bf k}|^{n_{\rm M}/2},
\end{equation}
where ${\bf f}({\bf k})$ is the Fourier transform of a $\delta$-correlated
vector field in three dimensions with Gaussian fluctuations, and
$n_{\rm M}=-3$ for a scale-invariant spectrum.
The degree of helicity is controlled by the parameter $\tilde\sigma$ and is
given by $\sigma=2\tilde\sigma/(1+\tilde\sigma^2)$. We assume $\rho=\rho_0=\const$ and $\uu=0$ initially, so the plasma is
at rest and the flow is generated from the magnetic field entirely by the Lorentz force.

The full system of equations was derived in Ref.~\cite{Brandenburg:1996fc} starting
from the general relativistic equations in an expanding
universe for a flat space-time. We use the ultrarelativistic equation of state where the pressure is given
by $p=\rho/3$ and the sound speed is $\cs=1/\sqrt{3}$. We assume the bulk velocity $\uu$ to be subrelativistic, so the
equations reduce to the usual MHD equations that have frequently
been used in the literature \cite{Christensson:2000sp,Banerjee:2004}, except that there are additional $4/3$ factors and some extra terms:
\begin{eqnarray}
{\partial\ln\rho\over\partial t}&=&-\frac{4}{3}\left(\nab\cdot\uu+\uu\cdot\nab\ln\rho\right)
+{1\over\rho}\left[\uu\cdot(\JJ\times\BB)+\eta\JJ^2\right] \\
{\partial\uu\over\partial t}&=&-\uu\cdot\nab\uu
+{\uu\over3}\left(\nab\cdot\uu+\uu\cdot\nab\ln\rho\right)
-{\uu\over\rho}\left[\uu\cdot(\JJ\times\BB)+\eta\JJ^2\right]\\
&&-{1\over4}\nab\ln\rho
+{3\over4\rho}\JJ\times\BB+{2\over\rho}\nab\cdot\left(\rho\nu\SSSS\right)\\
{\partial\BB\over\partial t}&=&\nabla\times(\uu\times\BB-\eta\JJ)
\label{dAdt}
\end{eqnarray}
where ${\sf S}_{ij}=\half(u_{i,j}+u_{j,i})-\onethird\delta_{ij}\nab\cdot\uu$
is the rate-of-strain tensor,
$\nu$ is the viscosity, and $\eta$ is the magnetic diffusivity. The differences compared to the standard MHD equations used in
Refs.~\cite{Christensson:2000sp,Banerjee:2004} are minor:
the kinetic energy would be overestimated by a factor of 4/3,
but the magnetic energy is about the same as for the full set
of equations. These differences will be discussed in a separate publication.

Our computational domain has a size $L$, so the smallest wavenumber is
given by $k_0=2\pi/L$.
We adopt a fully or partially helical magnetic field initially with a
scale-invariant spectrum $P_{\rm M}(k) \propto k^{-3}$ initially. We define the Alfv\'en time based on the value of $B_0$ with $B_0/\rho_0^{1/2}\cs=0.3$ as $\tauA=(B_0 k_0)^{-1}$ and measure conformal time in units of $\tauA$. The rms velocity is $\urms\approx0.03$ and $\rho_0=1$.

We use the {\sc Pencil Code} \cite{pencil}, which is a public domain code
for solving partial differential equations on massively parallel machines.
We use a spatial resolution of $1152^3$ meshpoints and set $\nu=\eta=
5\times10^{-6}\cs/k_1$, so the Reynolds number $\urms\xiM/\eta$ is
about $10^4$.
This is large enough so that the precise values of $\nu$ and $\eta$
are not expected to play any role.
Moreover, their ratio is taken to be unity, although simulations show that
the results still depend on this ratio \cite{Bra14}.

\begin{figure}[t!]\begin{center}
\includegraphics[width=.7\columnwidth]{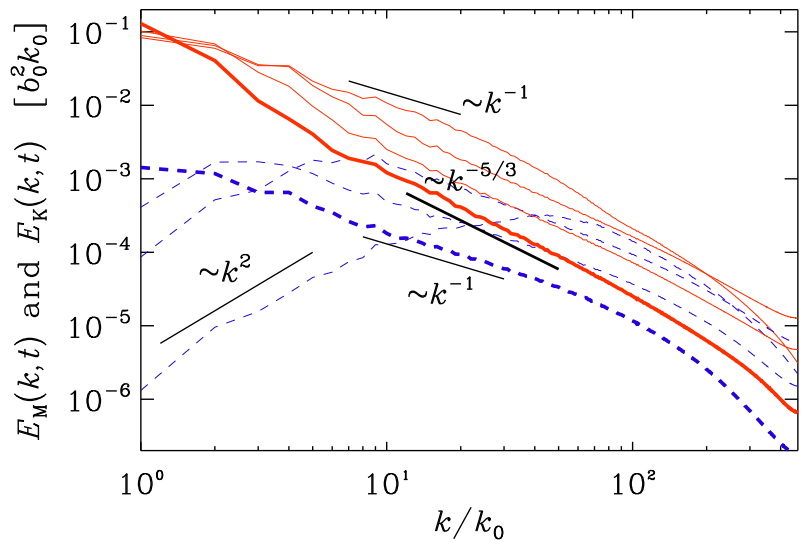}
\end{center}\caption[]{
Magnetic (red solid lines) and kinetic (blue dashed lines) energy spectra for $\tilde\sigma=\sigma=1$ at times $t/\tauA=0.03$, $0.3$, $1.2$, and $5$.
The last time is shown in boldface.
For orientation, the $k^2$, $k^{-1}$, and $k^{-5/3}$ slopes are indicated.
}\label{pkt1152_Mm1Kol1152_sig1_all}\end{figure}

\begin{figure}[t!]\begin{center}
\includegraphics[width=.7\columnwidth]{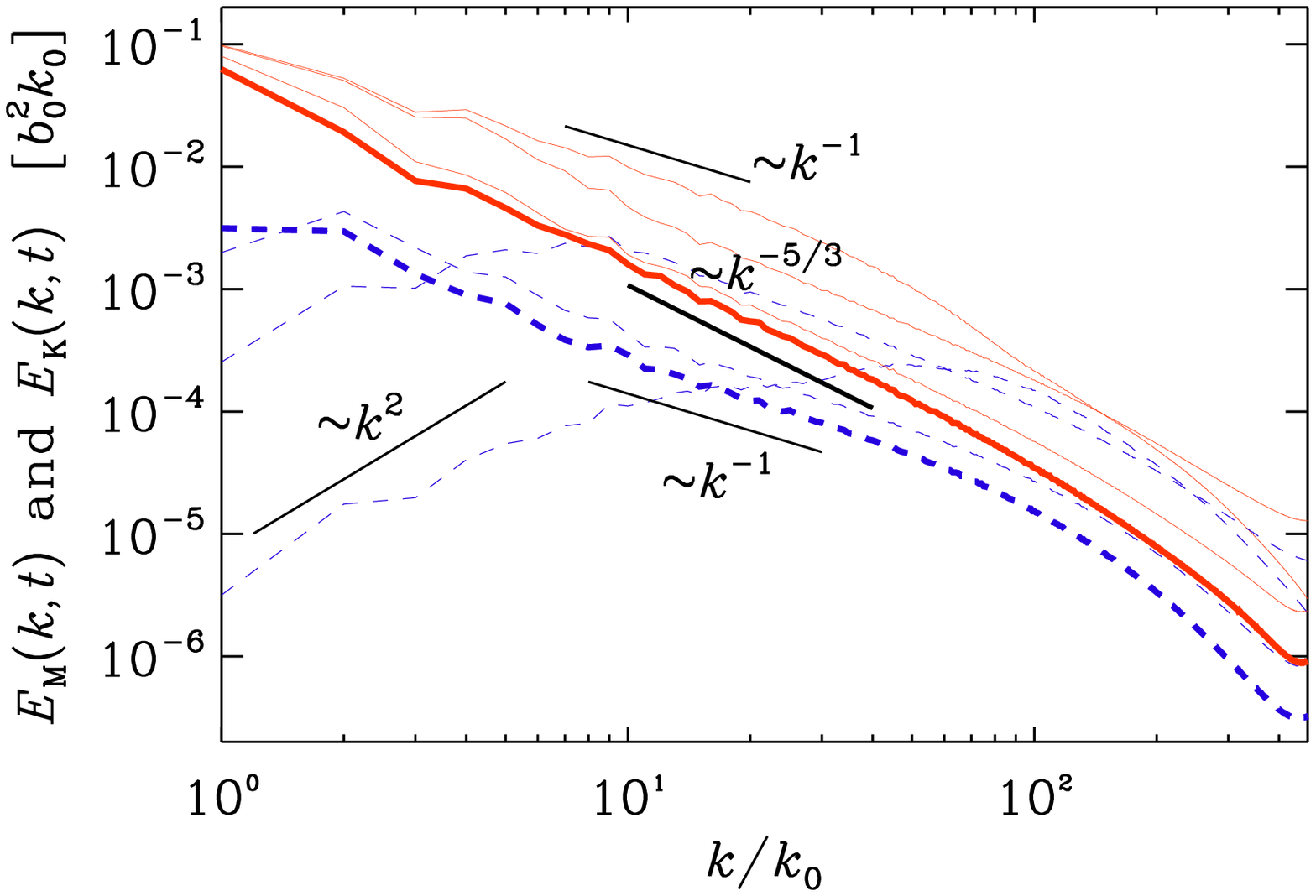}
\end{center}\caption[]{Same as \Fig{pkt1152_Mm1Kol1152_sig1_all}, but for $\tilde\sigma=0.03$
($\sigma\approx0.06$).
}\label{pkt1152_Mm1Kol1152_sig003c_all}\end{figure}

As time goes on, the inflationary helical magnetic field generates
helical fluid motions that are characterized by a white noise
spectrum, $E_K(k) \propto k^2$, at large, super-horizon scales,
typical of causal fields; see \Fig{pkt1152_Mm1Kol1152_sig1_all}. The magnetic field gets tangled by the resulting velocity field.
This leads to a forward cascade of magnetic energy with
a slope corresponding to a $k^{-5/3}$ Kolmogorov spectrum at larger
wavenumbers.
This result is virtually independent of the initial magnetic helicity, as can be seen from the corresponding results for $\tilde\sigma=0.03$;
see \Fig{pkt1152_Mm1Kol1152_sig003c_all}.

\begin{figure}[t!]\begin{center}
\includegraphics[width=.7\columnwidth]{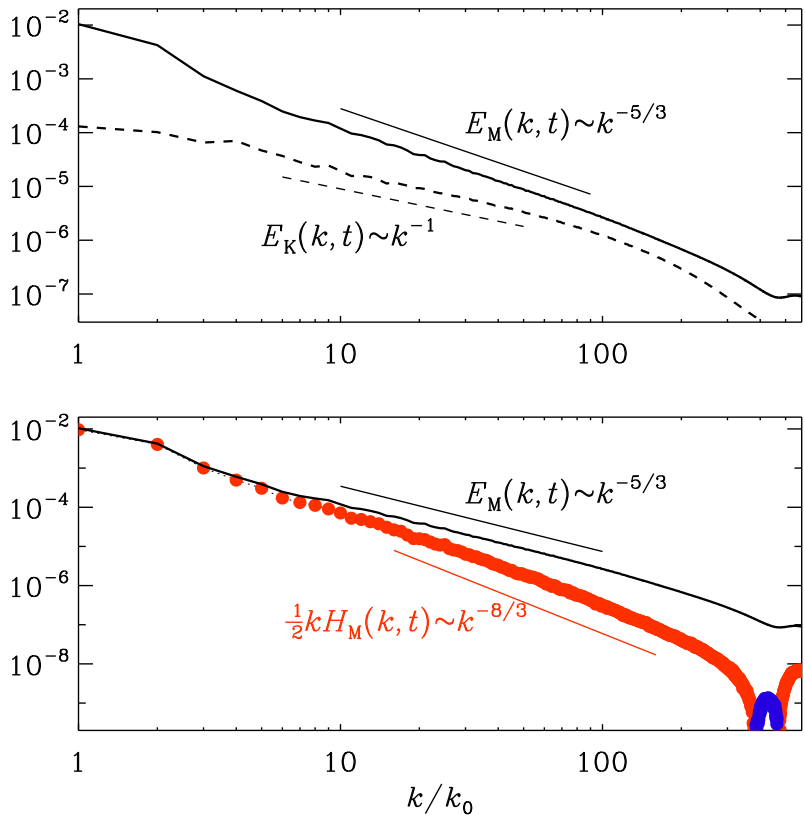}
\end{center}\caption[]{
Magnetic and kinetic energy spectra (upper panel) as well as
magnetic energy and scaled magnetic helicity spectra (lower panel)
for $\sigma=1$ in units of $b_0^2k_0$.
Positive (negative) values of $H_{\rm M}(k)$ are indicated
by red (blue) symbols.
}\label{pkt1152_Mm1Kol1152_sig1}\end{figure}

At the last time of the simulation, the kinetic energy spectrum has
approached the magnetic spectrum at small scales,
although $E_{\rm K}(k,t)<E_{\rm M}(k,t)$ in all cases.
Furthermore, from intermediate $k$ values onward, we find $E_K\sim k^{-1}$
and the initial $k^2$ subrange has completely disappeared; see the
upper panel of \Fig{pkt1152_Mm1Kol1152_sig1}.
We also compare with magnetic helicity spectra, normalized by $k/2$,
which allows us to see how close the realizability condition,
${1\over2}k |H_{\rm M}(k)|\leq E_{\rm M}(k)$, is to saturation
on large scales; see the lower panel of \Fig{pkt1152_Mm1Kol1152_sig1}.
It turns out that at large scales, where the magnetic field has
not yet been affected by the flow, the magnetic field is still nearly
fully helical.
However, for $k/k_1>10$, this is no longer the case and we see
that $kH_{\rm M}(k)\sim k^{-8/3}$, i.e., $H_{\rm M}(k)\sim k^{-11/3}$
at sufficiently late times.
This behavior has previously also been seen in forced turbulence
simulations \cite{Brandenburg:2005xc,Brandenburg:2008tj} and is a consequence of a forward cascade of current helicity, $k^2H_{\rm M}(k)$, also exhibiting a $k^{-5/3}$ spectrum.

\begin{figure}[t!]\begin{center}
\includegraphics[width=.7\columnwidth]{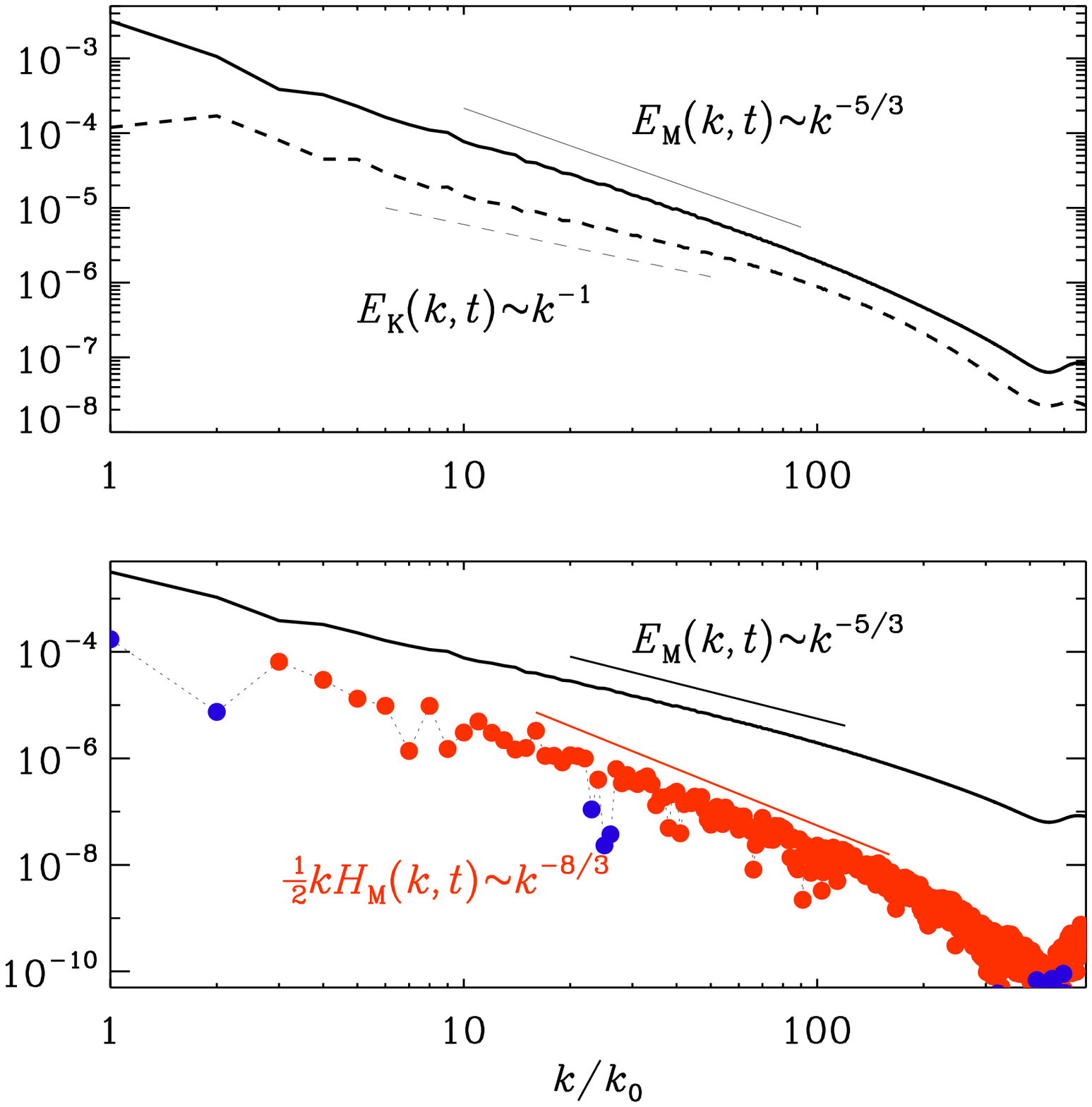}
\end{center}\caption[]{
Similar to \Fig{pkt1152_Mm1Kol1152_sig1}, but for $\tilde\sigma=0.03$
($\sigma\approx0.06$).
}\label{pkt1152_Mm1Kol1152u1_sig003c}\end{figure}

For $\tilde\sigma=0.03$, i.e., $\sigma\approx0.06$,
we find similar characteristics, except that now
${1\over2}k |H_{\rm M}(k)|\ll E_{\rm M}(k)$;
see \Fig{pkt1152_Mm1Kol1152u1_sig003c}.
There are now also more data points where $H_{\rm M}(k)$ is negative.
Nevertheless, we still see a clear $k^{-8/3}$ subrange in $kH_{\rm M}(k)$.

\begin{figure}[t!]\begin{center}
\includegraphics[width=.7\columnwidth]{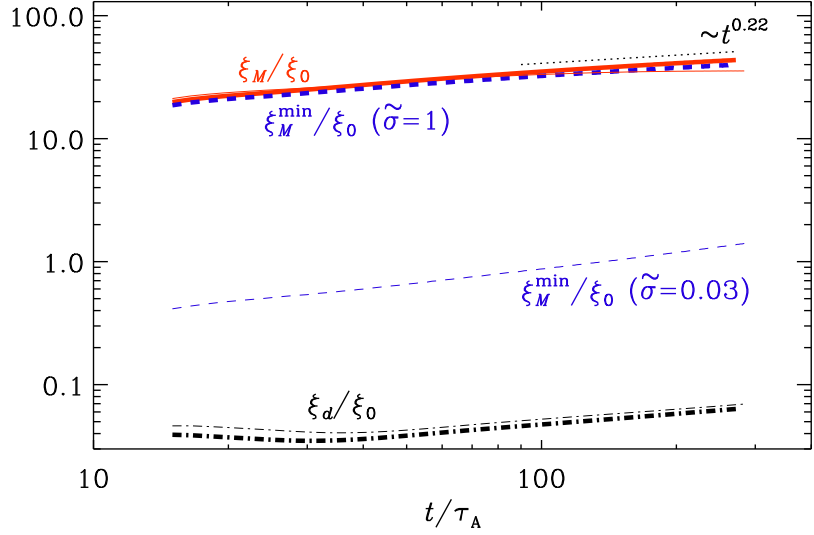}
\end{center}\caption[]{
Time evolution of $\xiM=\kM^{-1}$ and $\xiM^{\min}$,
as well as the Taylor microscale $\xi_d$.
The thin lines are for $\tilde\sigma=0.03$ and the thick ones for $\tilde\sigma=1$.
}\label{pcomp_kft_QCD_inflation}\end{figure}

In the present simulations, the initial magnetic energy spectrum
had significant energy at large scales, so $\xi_{\rm M}$ is already large
initially.
Thus, $\xi_{\rm M}$ can only grow owing to the decay of magnetic
energy at high wavenumbers.
This results in a temporal growth $\xi_{\rm M}\sim t^{0.2}$;
see \Fig{pcomp_kft_QCD_inflation}. In the case with $\tilde\sigma=0.03$, we find that $\xi_{\rm M}$ first increases
and then decreases by a certain amount.
This is different from the case of an initial $k^4$ subinertial range
energy spectrum with fractional helicity, where $\xi_{\rm M}$ is at first
growing proportional to $t^{1/2}$.
However, as has been demonstrated earlier \cite{Tevzadze:2012kk},
the growth of $\xi_{\rm M}$ speeds up when it reaches the value
$\xi_{\rm M}^{\min}\equiv|{\mathcal H}_{\rm M}|/2\xi_{\rm M}
{\mathcal E}_{\rm M}$, i.e., when the field has become fully helical
and thus $\sigma=1$; see Eq.~(\ref{kmin1}).
In the present case of an initial $k^{-1}$ energy spectrum,
$\xi_{\rm M}^{\min}$ also grows (see the dashed line in \Fig{pcomp_kft_QCD_inflation} for $\tilde\sigma=0.03$), but the growth
is too slow to become significant.
\begin{figure*}[t]\begin{center}
\includegraphics[width=.40\columnwidth]{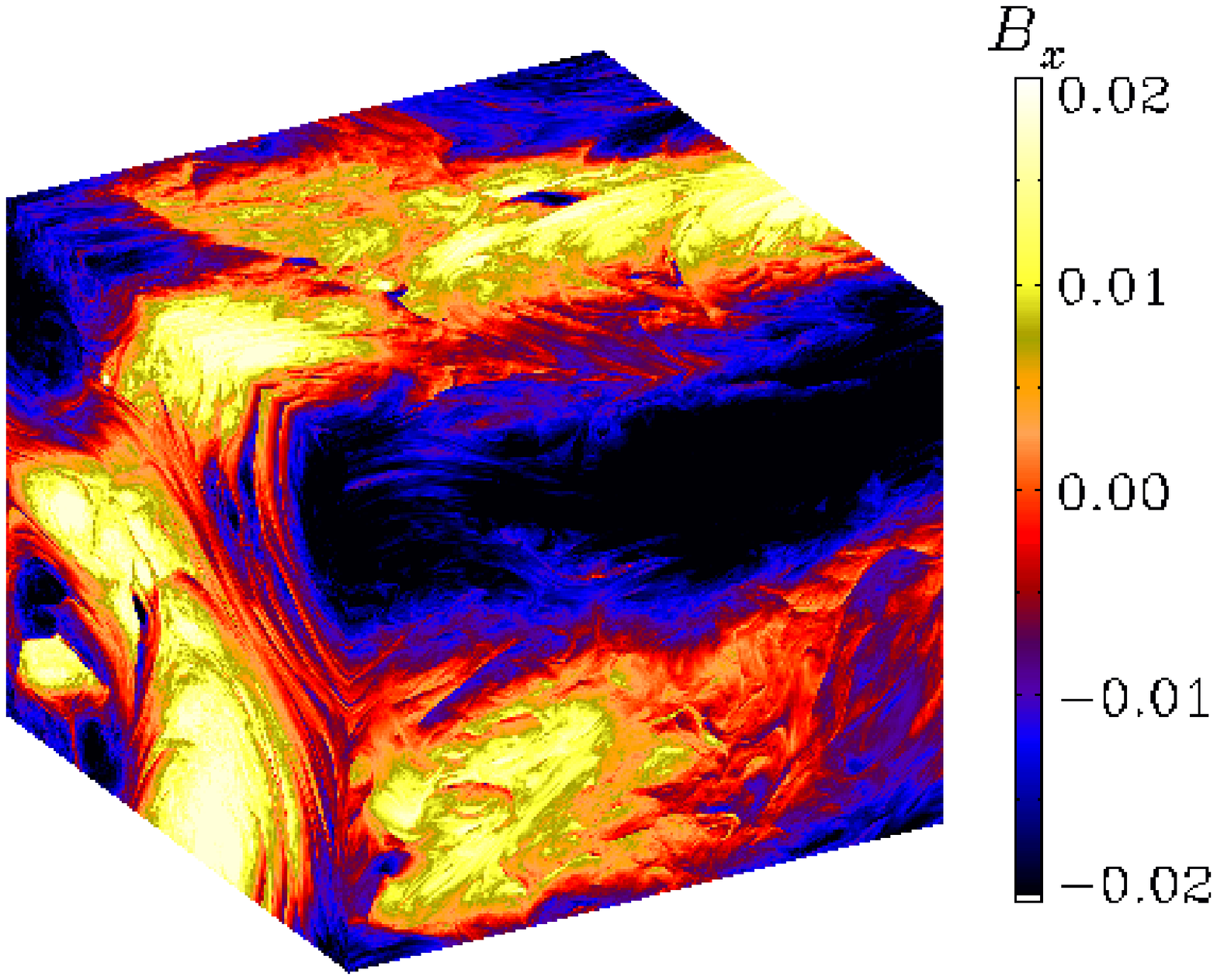}
\includegraphics[width=.40\columnwidth]{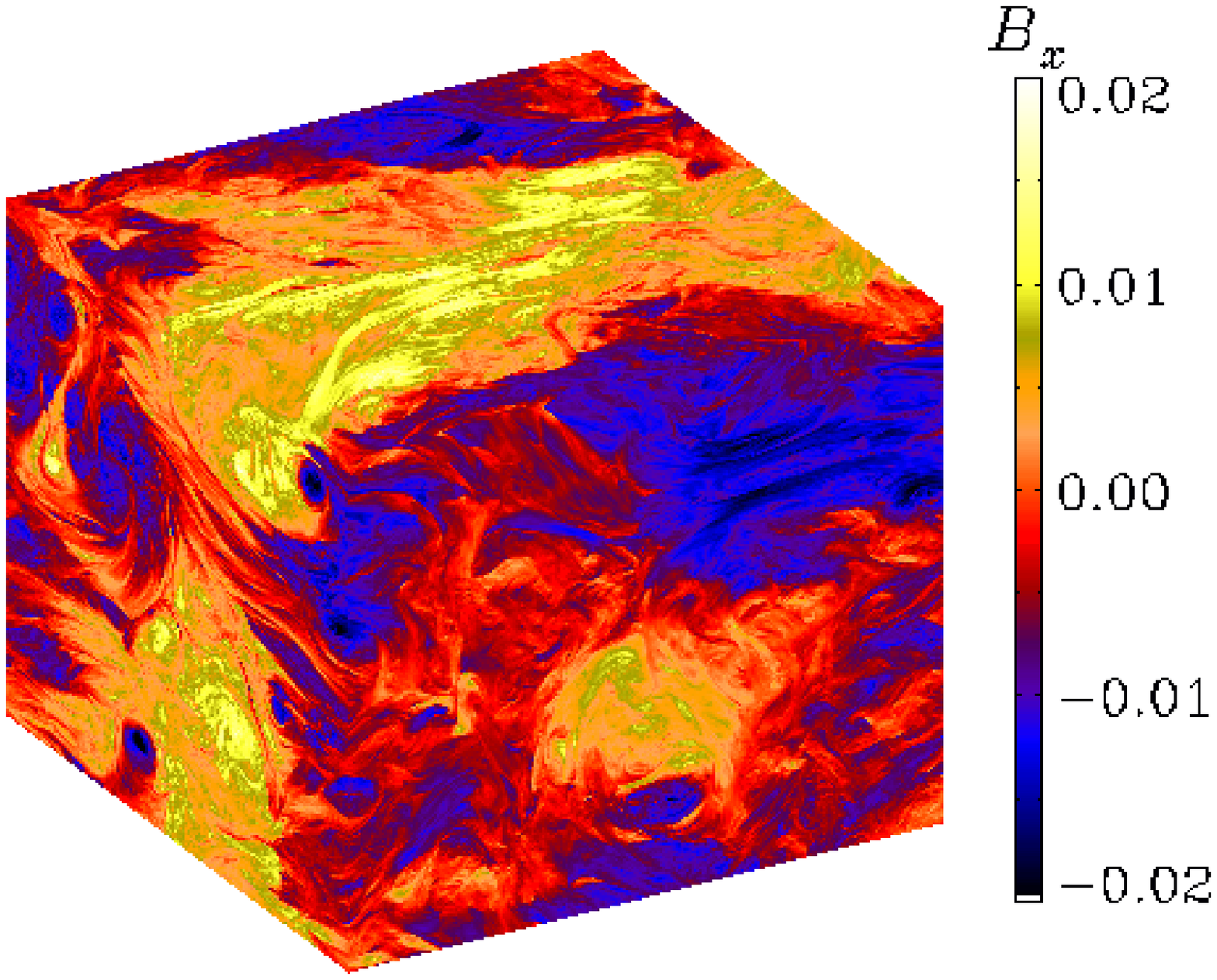}
\includegraphics[width=.40\columnwidth]{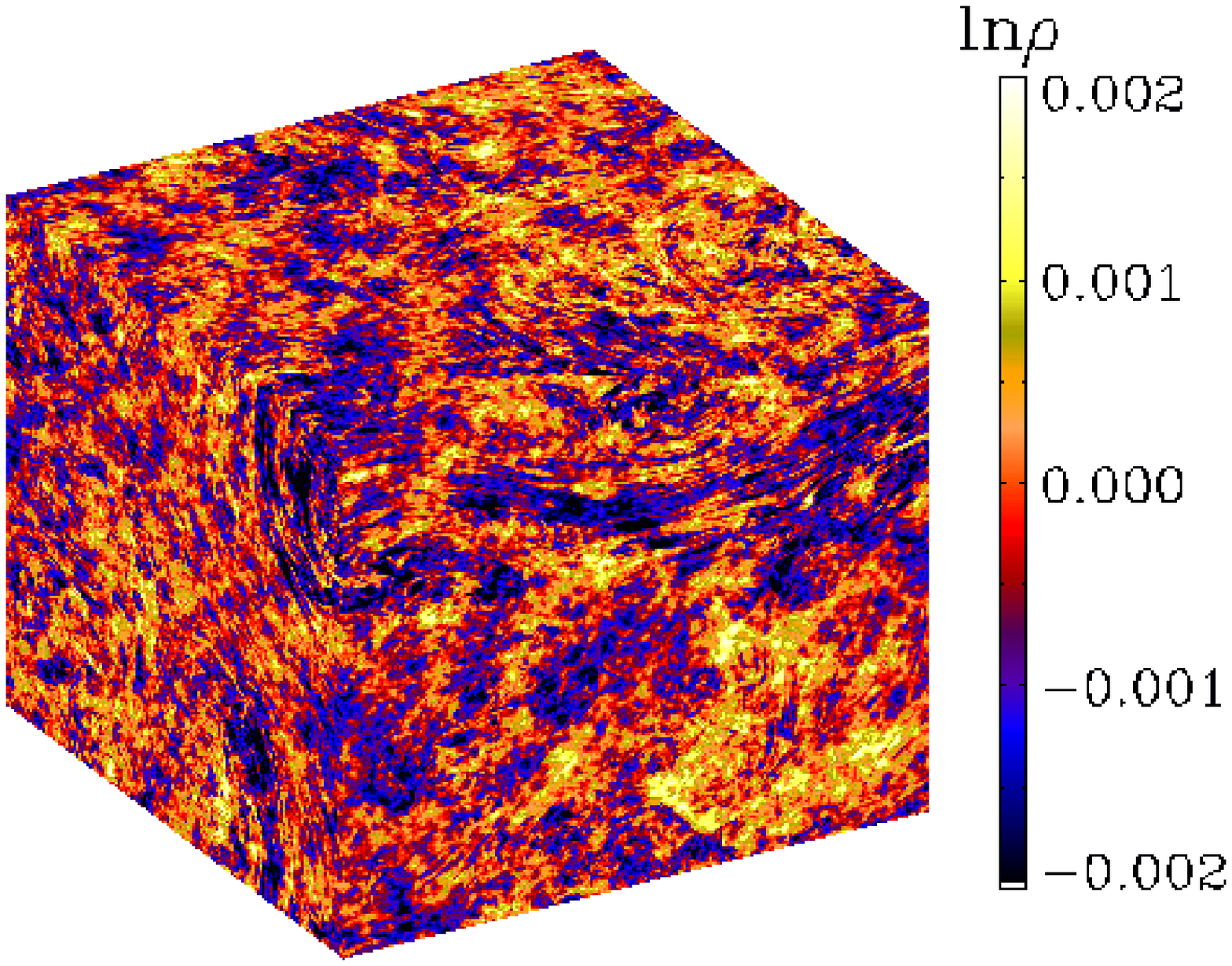}
\includegraphics[width=.40\columnwidth]{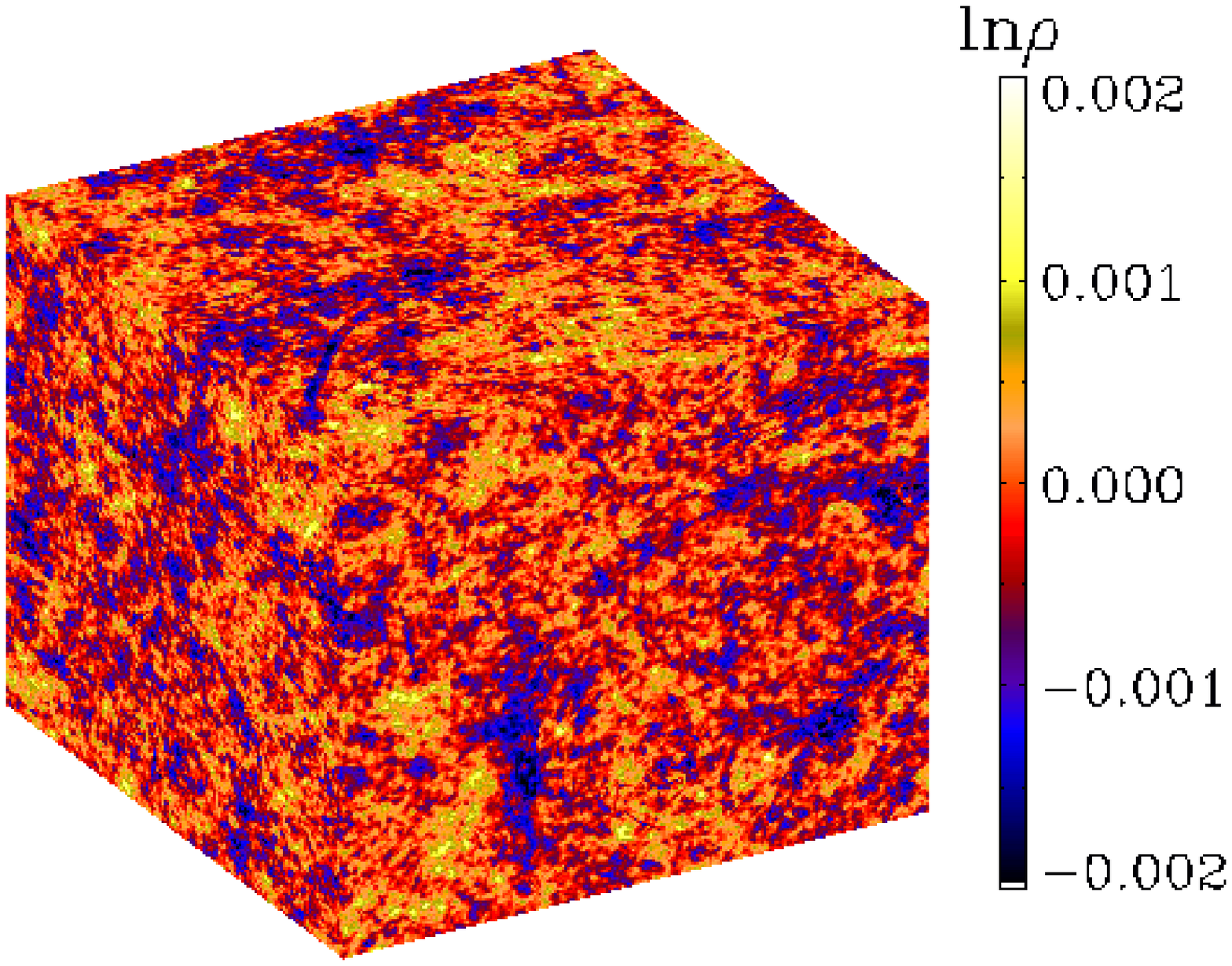}
\end{center}\caption[]{
Comparison of $B_x$ (upper row) and $\ln\rho$ (lower row) for $\tilde\sigma=1$ (left) and $\tilde\sigma=0.03$ (right).
}\label{bb1_sig003}\end{figure*}

Both for $\tilde\sigma=1$ and for $\tilde\sigma=0.03$, the magnetic field
has large-scale structure; see the upper panels of \Fig{bb1_sig003}.
In some locations, the field appears to have a folded structure
with sheets of alternating sign close together.
This feature is more pronounced in the case with $\sigma=1$ and
reminiscent of similar foldings in forced MHD simulations
at large magnetic Prandtl numbers, $\nu/\eta\gg1$ \cite{Schekochihin:2003bg}.
In both cases, the effect on the density perturbations is small,
but one still sees large-scale patches together with smaller scale
structures.

\section{Observational consequences \label{s:prospects}}

We have discussed scale-invariant helical magnetic fields generated
during inflation.
We have shown that in the case of a maximally helical magnetic field,
the ratio of magnetic helicity and energy densities gives a lower limit of the horizon scale at the beginning of inflation, i.e.,
\be
k_{\min}^{-1} \ge \xi_{\rm M} \ge \frac{|{\cal H}_{\rm M}|}
{2{\cal E}_{\rm M}} \,.
\ee

We have found numerically that, while the correlation length grows
substantially during the inverse cascade of a causal helical field,
this growth is strongly reduced for scale-invariant fields.
Consequently, also a fractional helicity $\sigma$ does not grow
significantly for scale-invariant helical magnetic fields.
Finally, we have argued that it
can in principle be measured in the CMB anisotropy and polarization.

The observable signatures of cosmological magnetic helicity and energy
density of the magnetic field on large scales have been studied in great
detail for the CMB through  Faraday rotation (see Ref.~\cite{Ade:2015cao}
and references therein), as
well as scalar (density), vector (vorticity), and tensor (gravitational
waves) perturbation modes (see Ref.~\cite{Ade:2015cva} and references therein), and for large-scale
structure (see Ref. \cite{Kahniashvili:2012dy} and references therein). For the helical magnetic fields the most relevant ones are vector
\cite{Kahniashvili:2005xe} and tensor modes \cite{Caprini:2003vc}
of perturbations, since the magnetic helicity is observable through the
CMB $B$-polarization that is sourced by the vector and tensor modes.
We neglect here the Faraday rotation effect, which is independent of
magnetic helicity \cite{Campanelli:2004pm,Kosowsky:2004zh}. We note, however, that, if there is a line-of-sight magnetic field
as well as intrinsic polarized emission
correlated with the magnetic field, there is in principle
the possibility that helical fields can produce a correlation or
anti-correlation (depending on the sign of helicity) with the rotation
measure \cite{Brandenburg:2014zia}.
However, this effect could be negligible for inflation-generated
fields where $k_{\min}$ is small.

The vector and tensor mode sources are given by the convolution of
the magnetic fields (see \cite{Kahniashvili:2005xe} for the vector mode
source and \cite{Caprini:2003vc} for the tensor mode source, respectively),
and as a result, even if the magnetic field is non-vanishing only for
$k_{\min} <k<k_{\max}$, the sources are finite also for $k<k_{\min}$.
In fact, the infrared white noise amplitude of a typical component of
the magnetic energy-momentum tensor is
\begin{equation}
\lim_{\bk\ra 0}\tau_{\mu\nu} (\bk) \simeq {\cal E}_{\rm M} \,.
\end{equation}
The helical structure of the source is reflected in the spectral form of
the induced perturbations (vorticity and gravitational waves) that have
now non-vanishing parity odd correlators (in the case of the tensor mode,
magnetic helicity leads to a net circular polarization of gravitational waves) \cite{Kahniashvili:2008er}.

We define the ratio between the antisymmetric and symmetric parts for
vector and tensor modes as ${\mathcal P}_V$ and ${\mathcal P}_T$.
In the absence of helical magnetic fields (or other parity odd sources)
these quantities vanish.
Under realistic conditions, the symmetric and antisymmetric parts of
the spectra (for both vector and tensor modes) scale differently
during the evolution of the universe.
We also expect that for a helical source of sufficiently long duration,
the induced fluctuations, vorticity or gravitational waves, become
maximally helical if the source is maximally helical. Therefore, at late times, ${\mathcal P}_V(k) \rightarrow 1$
and ${\mathcal P}_T(k) \rightarrow 1$ for $k \geq k_H$,
where $k_H$ denotes the mode that enters the horizon at time $t$
such that $k_H \sim 1/(at)$.
The observation of such a parity-odd component of vorticity and gravitational waves from helical magnetic fields in principle allows us
to determine both magnetic energy and magnetic helicity densities.

Existing CMB data on the off-diagonal cross-correlations between
temperature and $B$-polarization \cite{Bennett:2012zja,Hinshaw:2013,
Larson:2010gs} limit the magnetic helicity to roughly
10\,nG$^2\Gpc$ \cite{Kahniashvili:2014dfa}.
Assuming that the nG limit on magnetic fields~\cite{Ade:2015cva},
$2{\cal E}_{\rm M} \lsim 0.1$nG$^2$, is close to a future detected value,
we obtain
\be
\left(\frac{|{\cal H}_{\rm M}|}{2{\cal E}_{\rm M}}\right)_{\rm opt}
\sim100\Gpc \,.
\ee
This is still about 20 times larger than the Hubble scale. Hence, if we were soon to detect a magnetic helicity and energy density close
to their present limits, this would represent a very exciting result.
However, if we assume a scale-invariant spectrum with a field strength close to the lower limit of $B\gsim 10^{-16}\G$, we obtain
\be
\left(\frac{|{\cal H}_{\rm M}|}{2{\cal E}_{\rm M}}\right)_{\rm true}
\leq10^{16}\Gpc\,.
\ee

\section{Conclusions \label{s:con}}

In this work we have shown that a detection of nearly maximal helicity,
$|{\cal H}_{\rm M}|\simeq {\cal H}_{\rm M}^{\max}$,
can be used to limit the horizon scale at the beginning of inflation,
\be
\frac{2{\cal E}_{\rm M}}{|{\cal H}_{\rm M}|} \gsim k_{\min} \geq |t|_{\rm in}^{-1} \,.
\ee
Here, $|t|_{\rm in} \sim (a_{\rm in}H_{\rm in})^{-1}$ is the comoving
horizon scale at the beginning of inflation.

This is the first time that a possibility of determining experimentally
the beginning of inflation has been proposed.
Of course, our method is only applicable if inflation generates not
only initial fluctuations for structure formation in the universe but
also a scale-invariant spectrum of helical magnetic fields.

\appendix
\section{Monochromatic Magnetic Field}
\label{Monochromatic}

The purpose of this appendix is to illustrate with a simple example
how the ratio (\ref{kmin1}) can be measured.
For this we consider a monochromatic wave
\bea
\BB &=& {\boldsymbol{\beta}}_1\cos({\bf k_*\cdot x}) +  {\boldsymbol{\beta}}_2\sin({\bf k_*\cdot x})\\
\AAA &=& {\boldsymbol{\alpha}}_1\sin({\bf k_*\cdot x}) -  {\boldsymbol{\alpha}}_2\cos({\bf k_*\cdot x})\\
\mbox{where}&&  \nonumber\\
 {\boldsymbol{\beta}}_i &=& {\bf k}_*\wedge {\boldsymbol{\alpha}}_i
\eea
For simplicity we assume that $\bf k_*$ and $\boldsymbol{\alpha}_i$ are
all orthogonal and $|{\boldsymbol{\alpha}}_i|=\alpha$.
The energy density of this field is ${\cal E}_{\rm M} =\frac{1}{2}\langle
B^2\rangle = k_*^2(\alpha_1^2+\alpha_2^2)/4 =k_*^2\alpha^2/2$. The
helicity is $|{\cal H}_{\rm M}| = |- {\boldsymbol{\alpha}}_1\cdot
{\boldsymbol{\beta}}_2 + {\boldsymbol{\alpha}}_2\cdot
{\boldsymbol{\beta}}_1|/2 = k_*\alpha^2$.
Hence, for this (maximally helical) case
\be
\frac{|{\cal H}_{\rm M}|}{2{\cal E}_{\rm M}} =k_*^{-1} \,.
\ee
This example is of course not very realistic as an average over many
wavelengths is required. In the cosmological context we perform ensemble
averages which, perhaps, are best compared with averaging ${\cal E}_{\rm M}$
and ${\cal H}_{\rm M}$ over many phases.

\acknowledgments It is our pleasure to thank Leonardo Campanelli,
George Lavrelashvili, Arthur Kosowsky, Kerstin Kunze, Sayan Mandal, and Tanmay
Vachaspati for useful discussions.
Support through the NSF Astrophysics and Astronomy Grant Program
(grants 1615940 \& 1615100),
the Research Council of Norway (FRINATEK grant 231444),
the Swiss NSF SCOPES (grant IZ7370-152581), and the Georgian Shota Rustaveli
NSF (grant FR/264/6-350/14) are gratefully acknowledged. TK acknowledges the International Center for Theoretical Physics (ICTP) Senior Associate Membership Program.

%%%
\newcommand{\yprd}[3]{, Phys. Rev. D {\bf #2}, #3 (#1).}
\newcommand{\yprl}[3]{, Phys. Rev. Lett. {\bf #2}, #3 (#1).}
\newcommand{\yapj}[3]{, Astrophys. J. {\bf #2}, #3 (#1).}
\newcommand{\yana}[3]{, Astron. Astrophys. {\bf #2}, #3 (#1).}
\newcommand{\ymn}[3]{, Mon. Not. Roy. Astron. Soc. {\bf #2}, #3 (#1).}

%%%

\vfill\bigskip\noindent\tiny\begin{verbatim}
$Header: /var/cvs/brandenb/tex/tina/inflation-hel/paper.tex,v 1.85 2017/11/22 20:08:06 brandenb Exp $
\end{verbatim}

\end{document}